\documentstyle[preprint,revtex]{aps}

\begin{document}
\draft
\begin{title}
An algorithm to extract the \\ spanning clusters and calculate\\
  conductivity in strip geometries
\end{title}

\font\institute=cmss10
\author{ {\bf F.~Babalievski} \\[2ex]
  {\institute HLRZ at the KFA J\"ulich, D-52425 J\"ulich, Germany} \\
  {\institute and} \\
  {\institute Inst. .General and Inorganic Chemistry, 1113 Sofia, Bulgaria} \\
  }
\receipt{8 May 1994}
\begin{abstract}
I present an improved algorithm
to solve the random resistor problem  using a transfer-matrix technique.
Preconditioning by spanning clusters extraction both
reduces the size of the conductance matrix
and yields faster execution times when compared to previous algorithms.
\end{abstract}
\narrowtext
\nonum\section{Introduction}
The method most frequently used for solving numerically  the random
resistor network (RRN) problem has changed over time surprisingly often:
 relaxation methods for solving Kirchhoff's
equations were adopted in the seventies, while the early 80's were the
time of the  random walk method; then  the transfer-matrix
(TM) approach \cite{DV} came into fashion; and next the
 node-elimination method came forth in the
90's \cite{Gingold,FraLob88}.
A ``Fourier acceleration'' method was also proposed
in mid-80's \cite{Batr}.
Renewed interest in direct methods to
solve the set of Kirchhoff equations arose after the paper of Edwards
{\it et al.\/} had been published in 1988 \cite{Sokal}:
standard algebraic multigrid(AMG) method generally used for solving
large linear sparse systems was applied.
In a recent paper \cite{GFF} the standard
Kirchhoff's set
was reduced by Green's function formulation of Kirchhof's laws.

The random walk method  is probably the worst among the methods listed
above.  Although in some applications the random walk method could be
more suitable than the others the main reason for its frequent use
appears to be the nice exposition given in Stauffer's famous
introductory book \cite{Sta}.
This method faces the same problem as many iterative methods for
solving the Kirchhoff's equations: its performance decreases rapidly
at the critical region of a metal-insulator-like phase transition -
s.c. critical slowing down(CSD). Random walkers diffuse anomalously
slow at criticality$(p=p_c)$, hence the diffusion constant
(i.e. the conductivity) estimations
require more computer time at $p=p_c$ than for $p > p_c$.
In the same way network size scaling at criticality leads to a faster
 increasing of numerical efforts than the number of resistors involved.
The origin of CSD is not so transparent when the Kirchhoff's equations are
iteratively solved. CSD amounts here in increasing the number of iteration
needed to reach a certain precision. Probably CSD stems from the fractal
geometry of the resistor network. Such geometry  leads to multifractal
 distribution of
voltage drops \cite{RoHa} across the net(if an external voltage is applied)
and in
this way reduces the speed of convergence in iterative solvings.

The AMG, transfer-matrix and node elimination methods are free of CSD
in a sense specific
for each method. The speculation that  AMG  method eliminates
 the critical slowing down completely (or almost completely)
relies mostly on numerical data \cite{Sokal} which is
far from exhaustive. The node elimination method calculates directly
  (without any voltage evaluations) the network conductance by applying
consecutively the star-triangle transformation in a way to reduce the number
 of sites in the network until one resistor is left.
 The computational effort is proportional to
the number of resistors so it is faster at $p=p_c$ than for any $p > p_c$.
In a similar manner the TM approach is faster at the critical point for a
given system size, but here computations scale with size in a
non-trivial way which will be discussed further.
It is important to note that a modification  of the
TM approach in order to evaluate the voltage drops distributions cite
is possible\cite{Bergm};
such modification is impossible within the node elimination method.

For all of these three methods, the "preconditioning" of the
system by extracting the connected (spanning cluster) or
bi-connected (percolation backbone) component could significantly improve
the method performance. For node elimination and AMG approaches such
extraction could be easily implemented.

In this paper I am concentrating on the TM approach, presenting a modified
algorithm, which allows preconditioning by extraction of spanning clusters.



A conceptually important feature of any
TM approach is that one does not have to consider the entire system
(or its states) at a time in order to calculate its physical
properties.  Typically, one only requires information about state $n$
in order to proceed to state $n+1$ and subsequently discards the
information about state $n$. In contrast, the known  ways
\cite{Sokal,Tarjan,HJH84,RxHn}
to
extract the backbone requires that the percolation structure is kept
in computer memory in its entirety.

I present a TM algorithm which is improved in comparison to the
previous TM formulations in two ways.
It has inherited the
important feature of ``voltage-source book-keeping'' from
an earlier modification of the ``canonical'' TM approach made by the
author \cite{ZfB91} for application to quasicrystalline and random
lattices. This feature makes possible a better utilization of the dilute
structure of the random networks. Second, the system is preconditioned
by spanning clusters extraction. The specific way of extraction reduces
significantly the memory requirements, which otherwise are very restrictive.

\section{The transfer-matrix approach}

The TM approach to the numerical solution of the RRN problem has been
presented first by Derrida \& Vannimenus in 1982 \cite{DV} and has
been elaborated subsequently by several groups
\cite{HJH3d,JSP84,HJH2d}.  Characteristic for the TM approach in two
dimensions (2D) is the use of infinitely long strips of finite width
$L$ cut from the resistor network and, analogously, in three
dimensions (3D) the use of ``bars.''  The similarity to the
transfer-matrix method of the statistical mechanics of spin systems
consists of the introduction of a matrix $A(M)$ which represents the
properties (in the RRN problem the conductivities --- see below) of
the semi-infinite strip between $-\infty$ and strip slice $M$. As,
e.g., in \cite{DV}, a strip slice of a resistor network on the square
lattice may consist of the vertical resistors in column $M$ and the
horizontal resistors which connect columns $M$ and $M-1$. Knowledge of
the conductivity matrix $A(M)$ and the resistor configuration in slice
$M+1$ is sufficient to calculate $A(M+1)$ for the next slice.
Iteration for all subsequent slices finally obtains the conductivity
of the whole strip. In the case of a resistor network of unit
resistors and insulators the long edges of the strip are thought of as
electrodes and the resistors in the upper and lower layers are taken
to have zero resistance.

If the resistor network has a fractal structure its conductance tends
to zero as the system size increases --- in analogy to its mass density
decreasing to zero. More quantitatively, the infinite spanning cluster
at the percolation threshold $p_c$ is a fractal for which finite-size
scaling theory shows\cite{FSS} that its conductivity should scale with
the system size $L$ as $L^{-t/\nu}$, where $\nu$ is the percolation
correlation-length exponent and $t$ is the percolation transport
exponent.

The TM approach has been used first for obtaining precise estimates of $t$
for percolation on the square and cubic lattices \cite{HJH3d,HJH2d}.
In Refs. \cite{HJH3d,HJH2d} the matrix $A(M)$ is updated  after
the addition of every single resistor (the program is published in
\cite{JSP84})
instead of using matrix
equations as in \cite{DV}. Thus, the calculations are simplified and
accelerated.

In order to define the matrix $A(M)$, we attach voltage sources to the
open ends of the resistors at the right end column of the growing
semi-infinite strip.

The matrix $A(M)$ is defined by attaching to the sites of the current
right end column of the semi-infinite strip voltage sources $V_i$,
where $i$ labels the row position in the strip and thus assumes values
from $1$ to $L$, the width of the strip.  Since the voltage-current
relations in the network are linear the current from any selected
sources, say source $j$, is a linear function of the voltages $V_i$,
\begin{equation} \label{e:1}
I_j=\sum_i^L A_{ji}(M) V_i.
\end{equation}
The relation (\ref{e:1}) defines the matrix elements $A_{ij}(M)$ of the
TM $A(M)$. From now on I will suppress the argument $M$ when no
confusion can arise.

When a {\it horizontal\/} resistor $R$ is added to row $k$ then the
matrix $A$ changes to $A'$ with matrix elements
\begin{equation} \label{e:2}
A_{ij}'= A_{ij}- \frac{A_{ik}A_{kj}R}{1+A_{kk}R}.
\end{equation}
For infinite $R$, a case that we encounter in insolator-resistor
mixtures, Eq. (\ref{e:2}) simplifies to
\begin{equation} \label{e:3}
A_{ij}'= A_{ij}'- \frac{A_{ik}A_{kj}}{A_{kk}}.
\end{equation}

When we add a {\it vertical\/} resistor between two adjacent
sites $k$ and $l$ of a
new column then four matrix elements change,
\begin{equation} \label{e:4}
\left.
\begin{array}{rcl}
A_{kl}' & = & A_{kl}-1/R,  \\
A_{lk}' & = & A_{lk}-1/R,  \\
A_{kk}' & = & A_{kk}+1/R,  \\
A_{ll}' & = & A_{ll}+1/R.
\end{array}
\right.
\end{equation}

{}From Eq. (\ref{e:1}) it is clear that, in the limit $M \longrightarrow
\infty$, e.g., the difference $A(M)_{LL}-A(0)_{LL}$
tends to the transverse conductance of the strip of width $L$.  From
analysis of the conductivity scaling of strips with different $L$ one
obtains the conductivity scaling exponent. For the percolation cluster
at $p_c$ this exponent equals the ratio $t/\nu$.

The advantages of the TM approach have been described in the
pioneering works \cite{DV} and \cite{JSP84}.  Here, I would like to
point out to the reader its main drawback: the size of the matrix $A$ and
the computational effort grow very fast with the strip width $L$.

In particular, the size of the matrix grows as $L^2$ and $L^4$ for
2D and 3D respectively. If we consider a site percolation model
in 2D then
addition of every new column leads to adding an average of $p^2 (L-1)$
horizontal and $p^2 (L-2) +2 p$ vertical resistors. Taking the size of
the matrix $A$ into account, we find that Eq.
(\ref{e:2}) is applied $\propto L^3$ times whereas
only $\propto L$ operations are required for Eqs. (\ref{e:4}).
Thus, it is clear that for widths
larger than $10-15$ lattice spacings more than 90\% of the time is
spent on calculating the relations (\ref{e:2}).
In 3D the situation is even worse: the
upper bound for the computational efforts scales as $L^6$. But in fact
this bound is overestimated: Ref. \cite{JSP84} points out that the
computational effort scales as $L^4$ due to the fact that the
matrix $A$ is sparse, i.e., most of its elements equal zero.

In the next section, I will describe a modification of the TM approach
that overcomes these problems in parts.

\section{The modified algorithm}

The site-percolation case will be considered without loss of
generality.  The ``voltage-sources book-keeping'' procedure is
described in the next subsection.  Second(Sec.II.B.) comes the method
for extracting the spanning clusters and at last(Sec.III.C.) I present
the main steps in the complete algorithm.

\subsection{Conductivity calculations}
\label{s:conductivity}

Let us reconsider Eq. (\ref{e:1}) in the case of a general
resistor network.
Let some network nodes of an arbitrary resistor network
be connected with external voltage sources $V_i$).
Then $I_j$ [Eq. (\ref{e:1})] is the current from source $j$ and the
absolute values of the off-diagonal elements of matrix $A_{ij}, i\neq
j;$ represent the conductances between voltage sources $i$ and $j$.
The diagonal elements give the conductance between the respective
source and the ``ground'' --- the other sources set to zero voltage.
In fact, we can likewise interpret the matrix elements $A_{ij}$
introduced in Eq. (\ref{e:1}) in the previous section. The difference
is that the number of sources in the present case does not strictly
depend on the strip width and their connection to the right side end
is not mandatory (see Fig. 1 in Ref. \cite{ZfB91}).  Eqs.
(\ref{e:2}-\ref{e:4}) still apply in the present general case if we
allow in the general geometry the ``addition of a vertical resistor''
as connection with a resistor of two sites with voltage sources
attached to them and the ``addition of a horizontal resistor'' as
insertion of a resistor between a site and the voltage source
previously attached to this site.  In other words, ``adding a
horizontal resistor'' creates a new site and moves the voltage source
to it.

Based on these general concepts, we may formulate as algorithm the
conductivity
calculations Eqs. (\ref{e:2}-\ref{e:4}) for the resistor
strip case and the construction of the strip. The algorithm has three
main steps.

\begin{quote}

(i) Add a new site to the right end of the already existing
strip and attach a voltage
source to this site.

(ii) Find  the neighbors of this site among the sites already present and
connect them with resistors. Update the matrix  using Eq. (\ref{e:4}).
The algorithm should ensure that the neighbors have
their own voltage sources attached,

and (iii) if a site with its attached voltage source is located within
the bulk of the growing strip then detach the source to free it for
subsequent attachment to a new site on the growing edge of the strip.
To this end, we (a) insert an insulating ``resistor'' between the
network site and the previously attached source. Then (b) we update
the TM according to Eq. (\ref{e:3}). The information about the voltage
source index is kept on a stack, ready to be (c) used again when a
new site is added to the right end of the strip.
We have done nothing more than adding a bridge of infinite resistance
between two points of the network which does not alter its
conductivity properties and moved a voltage source.

\end{quote}

Since a regular lattice structure of the resistor bonds is not a
prerequisite for the conductivity calculation outlined above, an
algorithm based on the steps (i-iii) is useful for calculation of
percolative conductivities of {\it quasicrystalline} and {\it random}
lattices \cite{ZfB91,PA92}.

Since we have to only update the TM for lattice sites that are
actually connected to the strip by resistors of finite resistance and
since we always apply the simpler Eq. (\ref{e:3}) instead of Eq.
(\ref{e:2}), the outlined
procedure is already faster than the standard algorithm\cite{JSP84}.

The matrix size, i.e., the number of matrix elements,
instead of being $L^2$ is only $\approx p^2 L^2$ for the square lattice and
$\approx p^2 L^4$ instead of $L^4$ for the cubic lattice.
The scaling with $L$ of the matrix size is not altered but the
way of handling the voltage source numbers allow for a significantly
smaller prefactor
and (more important) it facilitates the system
preconditioning which does lead to improvement of memory and performance
scaling.

\subsection{Spanning clusters extraction}  

We achieve an improvement of the scaling of the matrix size as a
function of $L$ by the extraction of spanning clusters.  The spanning
clusters are defined as the percolating clusters which connect the top
and bottom edges of the strip or, respectively, the bottom and top
faces of the bar in 3D. At $p_c$ the spanning clusters in strip
geometries represent the incipient infinite percolation cluster.  Its
fractal dimension $d_f$ is $91/48$ in 2D and around $2.5$ in 3D
\cite{Sta}. If only voltage sources connected to the spanning clusters
contribute to the matrix size then this size should scale as $L^{2
  (d_f-1)}$, where the exponent $d_f-1$ reflects the system width
dependence of the scaling of the spanning cluster's sites found in a
$d-1$ dimensional cut \cite{Vicz}.  Thus, in 2D the matrix size scales
as $L^{1.79..}$ instead of $L^2$. In 3D the number of the matrix
elements is $\propto L^{3. ..}$ rather than $L^4$.

How does one extract the spanning clusters in strip geometries?
Several general algorithms exist to solve this problem
\cite{Tarjan,HJH84,RxHn}. However, they all require that the
percolation structure has been created beforehand and is stored in its
entirety leading to large computer memory requirements ( e.g. for a bar
$10^5 \times 100 \times 100$ one has to consider $10^9$ sites
and $10^8$ matrix elements).

I now
present an algorithm which partly resolves this problem which would
otherwise limit the strip length.

The method for extracting the spanning cluster is based on the
Hoshen-Kopelman algorithm \cite{HK,Sta} for cluster counting. As is
well-known this algorithm requires only
consecutive $d-1$ dimensional cuts of the
lattice to be kept during its lattice ``sweeping.'' Cluster
information is stored in one 1D array --- the array of cluster sizes
and pointers, sometimes denoted as the array of ``labels of labels (LOL)''
\cite{Sta}.  The index into this array represents the cluster labels
and its elements are either ``cluster roots'' --- then containing the
size of a specific cluster --- or pointers to these cluster roots ---
i.e., negative numbers whose absolute value is equal to the index of
the array element corresponding to the cluster root.  Moreover, the
cluster root element may be used to store other information about its
cluster, e.g., whether the cluster touches the upper or/and lower
layer of the strip.

Running the Hoshen-Kopelman algorithm requires that the percolation
structure be scanned twice in order to extract the spanning clusters.
These two runs are required because if we reach a site during the
first run we cannot decide if this site's cluster will eventually turn
out to span.  To avoid storage of the entire cluster in memory, we
perform the second sweep based on a repetition of the pseudo-random
number sequence that created the first sweep cluster.  During the
second sweep the LOL array is examined to decide which
cluster a site belongs to and whether this cluster spans. Only sites
belonging to spanning clusters enter the conductivity calculations.

Thus, instead of storing the percolation cluster structure itself, we
only store the LOL array. The key question for the
proposed algorithm is the size of the LOL array that has to be
retained in memory between the two Hoshen-Kopelman sweeps. To
keep its size small I apply a procedure to recycle unused
labels \cite{Bind,HWF}. The size of the resulting LOL
arrays turns out to be less than 0.5\% of the memory required
to retain the whole cluster structure in memory.

\subsection{ }

The full algorithm, including both the
spanning clusters extraction and the conductivity-calculation procedures,
may be summarized as follows:

\begin{enumerate}
\item scan the random structure with the Hoshen-Kopelman procedure
constructing the LOL array by a label recycling technique.
After the sweep, keep the LOL array in memory.
\item repeat the scan using {\it the same pseudo-random number sequence:}
\begin{quote}
after creation of a new site DO:

\begin{quote}
decide by comparing the new and the stored LOL array whether this site
belongs to a spanning cluster. If it does then the site enters the
TM conductivity calculations. These proceed according to step (i--iii)
as outlined  in the previous subsection.

\end{quote}

\end{quote}
\item When the second scan terminates calculate the
transverse conductance per unit length as,
\begin{equation} \label{e:5}
\Sigma_L=\frac{ A(M)_{LL} -A(M_0)_{LL}}{M-M_0},
\end{equation}
where I have used $M_0 = M/5$ to reduce boundary effects.

\end{enumerate}

\section{Performance scaling results}

I have developed the algorithm described in the preceding
section in conjunction with a study \cite{HZ96} on the
conductivity of several distinct percolation models and has not
only been applied to standard percolation.
%

I compare the proposed in this paper TM algorithm(the modified algorithm)
to the previously published \cite{JSP84,ZfB91} TM algorithms
(the standard algorithms). As a standard algorithm I used mostly the
algorithm proposed in \cite{ZfB91} which was described in Sec II A.
The code published in \cite{JSP84} was run with a technical improvement
(zero elements check in the most-inner loop) only to be seen that
the performance scaling is the same for both "standards". It is worth
to note that the performance scaling $\sim L^4$ in 3D,
reported in \cite{JSP84}, is an overestimation probably due to that
technical item.


In Fig.~1 I display the amount of computer time required for the
conductivity calculations by the modified and the standard algorithm
on two different percolation models, namely ordinary site
percolation
and one step bootstrap percolation \cite{HZ96}.
In the bootstrap percolation model\cite{Adler} one generates a site
configuration in several steps.  First, one randomly occupies a specific
small fraction of the lattice sites. Subsequently, one determines all
empty lattice sites with at least two occupied neighbors
and occupies these empty sites as well. The steps are repeated until no empty
sites with two occupied neighbors remain. If such procedure stops after its
first step I call it one step bootstrap percolation.
The percolation-transport and correlation-length exponents of the
one step bootstrap percolation model almost equal those of ordinary site
percolation \cite{HZ96}.
The computer time for these two models,
when running the standard algorithm in 3D, scaled in the same way -
even with the same
prefactor, so the averaged results are given on one curve (``3d'') on Fig.~1.
This coincidence encouraged using the data available\cite{HZ96}
in 2D for the comparison in the next paragraph.
( In 2D the two algorithms were applied to different models: the standard
algorithm to the
bootstrap model and the modified to ordinary percolation. )

As expected from the arguments in the preceding section, the modified
algorithm
 displays the better scaling properties throughout.
In 3D site percolation the computer time scales as $\sim L^{3.24}$ for
the standard algorithm,
whereas the modification needs time proportional to $L^{2.40}$ only.
Similarly, in 2D  we observe that the standard  requires time
$\sim L^{2.02}$ whereas $\sim L^{1.64}$ suffices for the modification.
One can see that these values are appreciably smaller than the respective
upper bonds given in Sec.3.2.

The errors in the above values, given by the LSQ fitting procedure
 were  smaller than the uncertainty coming from the eventual
correction-to-scaling terms.  A more careful analysis is needed,
but a reasonable value for the error should be within $0.1$ - $0.2$.
All the tests were made at the percolation threshold for the respective model.
The strips(bars) length was of the order $10^6$ in two dimensions and $10^5$
in 3D. Several strips were calculated for each width and model.
If one have
to compare with  statistics accumulated by means of other method
on square(cubic) samples,
a strip $10^6 \times 100 $ may correspond to several thousend
runs
on (say) $100 \sqrt{2} \times 100 \sqrt{2}$ sample.(In Ref.\cite{MonAlb}
was found numerically that at $p_c$ the average length of the
spanning clusters
is around $2.0 L$ and their number is $\cong (3/8) (M/L)$ where $M$ is the
strip length.)

The modified model was applied as well in studying properties of some
``percolation-generated'' fractals. After the extraction of spanning clusters
at the percolation thresholds one adds new sites on the cluster perimeter
in order some aerogell structures to be modelled \cite{stoll,nakan}.

In Fig.~2 I display timing results for such fractals
 (cf., \cite{HZ96}).
The data sets (I) and (II) correspond to
two ensembles with higher (I) and lower (II) fraction of additionally
occupied perimeter sites of the spanning clusters. The set (III)
has been taken for a model in which one considers
second nearest neighbors in the spanning clusters as connected.
As seen depending on the model the computer time scaling
may vary significantly.

\section{Conclusion}

The modified TM algorithm proposed in this work reduces significantly
the computational efforts required for obtaining the conductivity
scaling for fractal structures in 2D and especially in 3D.
In contrast to the $L^{3.2..}$ computer time requirements of conventional TM
algorithms in 3D the time requirements of the algorithm proposed in this work
scales approximately as $L^{2.4}$ for percolation clusters at $p_c$.

Extracting the percolation cluster backbone,  instead
of spanning clusters only, would ensure a further improvement of the
performance.

Probably the main disadvantage of the modification of the TM
approach described here is the complexity of the algorithm.
Therefore, I have made the program publicly available \cite{HWF}.

\nonum\section{Acknowledgments}

I would like to thank H.~J.~Herrmann and Many Particle Group at the
HLRZ for their hospitality. I am grateful to W.~Verm\"o{}hlen for his
help during my stay at the HLRZ.  S.~Melin and H.~Puhl helped me to use
the computational facilities at the HLRZ. I acknowledge H.~J.~Herrmann
for the comments on the preliminary version of the manuscript. Special
thanks to S.~Schwarzer who helped me a lot in improving the final text.

This work was  supported by the European Communities commission on science
Grant No.CIPA3511PL920176 and  by Bulgarian NSF Grant No.F-119/91.

\

{\bf \noindent Figure captions}

\bigskip
{\bf Fig.~1} Comparison between the performance with and
without extracting spanning clusters. (Time units are
Sun SPARC 10 workstation CPU seconds)

\bigskip
{\bf Fig.~2} Performance scaling for modified percolation models
where, after extraction of spanning clusters, additional loops were
closed: (I) by addition of new sites on the cluster perimeter; (II) by
adding sites as in (I) but with a lower density, and (III) by
increasing the connectivity range to second neighbors. (Time units are
Sun SPARC 10 workstation CPU seconds.)

%
\clearpage

\end{document}